\documentclass[preprint]{elsarticle}

\usepackage{graphicx}
\usepackage{color}
\usepackage[dvipsnames]{xcolor}
\usepackage{xfrac}
\usepackage{epsfig}
\usepackage{hyperref}
\usepackage[all]{hypcap}
\usepackage[letterpaper,bindingoffset=0.0in,left=1in,right=1in,top=1in,bottom=1in,footskip=.25in]{geometry}
\usepackage{xspace}
\usepackage{booktabs}
\usepackage{amsmath}
\usepackage{multirow}
\usepackage{subcaption}
\usepackage{csquotes}

\makeatletter
\def\ps@pprintTitle{%
 \let\@oddhead\@empty
 \let\@evenhead\@empty
 \def\@oddfoot{}%
 \let\@evenfoot\@oddfoot}
\makeatother

\usepackage{xpatch}
\xpatchcmd{\MaketitleBox}{\hrule}{}{}{}
\xpatchcmd{\MaketitleBox}{\hrule}{}{}{}
\xpatchcmd{\pprintMaketitle}{\hrule}{}{}{}
\xpatchcmd{\pprintMaketitle}{\hrule}{}{}{}

\usepackage{etoolbox}
\patchcmd{\abstract}{Abstract}{\vspace{-\baselineskip}}{}{}

\newcommand{\csi}{CsI{[Na]}\xspace}
\newcommand{\cevns}{\protect{CE$\nu$NS}\xspace}
\newcommand{\etal}{\emph{et al.}\xspace}
\newcommand{\us}{\protect{\ensuremath{\mu\text{s}}}\xspace}
\newcommand{\args}[1]{\!\left( #1 \right)}

\newcommand{\nue}{\protect{\ensuremath{\nu_e}}\xspace}
\newcommand{\numu}{\protect{\ensuremath{\nu_\mu}}\xspace}

\newcommand{\numubar}{\protect{\ensuremath{\overline{\nu}_\mu}}\xspace}

\newenvironment{keeptogether}{\par\noindent\begin{minipage}{\linewidth}}{\end{minipage}\par}

\usepackage{listings}
\definecolor{mygreen}{rgb}{0,0.6,0}
\definecolor{mygray}{rgb}{0.5,0.5,0.5}
\definecolor{mymauve}{rgb}{0.58,0,0.82}
\newcommand\YAMLcolonstyle{\color{red}\mdseries}
\newcommand\YAMLkeystyle{\color{black}\bfseries}
\newcommand\YAMLvaluestyle{\color{blue}\mdseries}

\makeatletter

\newcommand\language@yaml{yaml}

\expandafter\expandafter\expandafter\lstdefinelanguage
\expandafter{\language@yaml}
{
  keywords={true,false,null,y,n},
  keywordstyle=\color{darkgray}\bfseries,
  basicstyle=\YAMLkeystyle,                                 
  sensitive=false,
  comment=[l]{\#},
  morecomment=[s]{/*}{*/},
  commentstyle=\color{purple}\ttfamily,
  stringstyle=\YAMLvaluestyle\ttfamily,
  moredelim=[l][\color{orange}]{\&},
  moredelim=[l][\color{magenta}]{*},
  moredelim=**[il][\YAMLcolonstyle{:}\YAMLvaluestyle]{:},   
  morestring=[b]',
  morestring=[b]",
  literate =    {---}{{\ProcessThreeDashes}}3
                {>}{{\textcolor{red}\textgreater}}1     
                {|}{{\textcolor{red}\textbar}}1 
                {\ -\ }{{\mdseries\ -\ }}3,
}

\lst@AddToHook{EveryLine}{\ifx\lst@language\language@yaml\YAMLkeystyle\fi}
\makeatother

\newcommand\ProcessThreeDashes{\llap{\color{cyan}\mdseries-{-}-}}

\begin{document}

\title{COHERENT Collaboration data release from the first observation of coherent elastic neutrino-nucleus scattering}
\address[kurchatov]{Institute for Theoretical and Experimental Physics named by A.I. Alikhanov of National Research Centre ``Kurchatov Institute'', Moscow, 117218, Russian Federation}
\address[mephi]{National Research Nuclear University MEPhI (Moscow Engineering Physics Institute), Moscow, 115409, Russian Federation}
\address[iu]{Department of Physics, Indiana University, Bloomington, IN, 47405, USA}
\address[tunl]{Triangle Universities Nuclear Laboratory, Durham, North Carolina, 27708, USA}
\address[duke]{Department of Physics, Duke University, Durham, NC 27708, USA}
\address[ut]{Department of Physics and Astronomy, University of Tennessee, Knoxville, TN 37996, USA}
\address[ornl]{Oak Ridge National Laboratory, Oak Ridge, TN 37831, USA}
\address[nccu]{Department of Mathematics and Physics, North Carolina Central University, Durham, NC, 27707, USA}
\address[sandia]{Sandia National Laboratories, Livermore, CA 94550, USA}
\address[efi]{Enrico Fermi Institute, University of Chicago, Chicago, IL 60637, USA}
\address[kicp]{Kavli Institute for Cosmological Physics, University of Chicago, Chicago, IL 60637, USA}
\address[uofcphys]{Department of Physics, University of Chicago, Chicago, IL 60637, USA}
\address[lbnl]{Lawrence Berkeley National Laboratory, Berkeley, CA 94720, USA}
\address[nmsu]{Department of Physics, New Mexico State University, Las Cruces, NM 88003, USA}
\address[lanl]{Los Alamos National Laboratory, Los Alamos, NM, USA, 87545, USA}
\address[uwcenpa]{Department of Physics and Center for Experimental Nuclear Physics and Astrophysics,\\ University of Washington, Seattle, WA 98195, USA}

\address[ncsu]{Physics Department, North Carolina State University, Raleigh, NC 27695, USA}

\address[pnnl]{Pacific Northwest National Laboratory, Richland, WA 99352, USA}

\address[itep]{Moscow Institute of Physics and Technology, Dolgoprudny, Moscow Region 141700, Russian Federation}

\address[cmu]{Carnegie Mellon University, Pittsburgh, PA 15213, USA}
\address[ufl]{Department of Physics, University of Florida, Gainesville, FL 32611, USA}

\address[laurentian]{Department of Physics, Laurentian University, Sudbury, Ontario P3E 2C6, Canada}
\address[kaist]{Department of Physics at Korea Advanced Institute of Science and Technology (KAIST)
and Center for Axion and Precision Physics Research (CAPP) at Institute for Basic Science (IBS), Daejeon, 34141, Republic of Korea}

\author[kurchatov,mephi]{D.~Akimov}

\author[iu]{J.B.~Albert}

\author[tunl,duke]{P.~An}

\author[tunl,duke]{C.~Awe}

\author[tunl,duke]{P.S.~Barbeau}


\author[ut]{B.~Becker}

\author[kurchatov,mephi]{V.~Belov}

\author[ornl]{M.A.~Blackston}

\author[mephi]{A.~Bolozdynya}

\author[nccu,tunl]{A.~Brown}

\author[kurchatov,mephi]{A.~Burenkov}

\author[sandia]{B.~Cabrera-Palmer}

\author[duke]{M.~Cervantes}

\author[efi,kicp,uofcphys]{J.I.~Collar}

\author[lbnl]{R.J.~Cooper}

\author[nmsu,lanl]{R.L.~Cooper}

\author[uwcenpa]{C.~Cuesta\fnref{clara}}
\fntext[clara]{Presently at Centro de Investigaciones Energ\'eticas, Medioambientales y Technol\'ogicas (CIEMAT), Madrid 28040, Spain}

\author[ut]{J.~Daughhetee}

\author[ornl]{D.J.~Dean}

\author[iu]{M.~del~Valle~Coello}

\author[uwcenpa]{J.~Detwiler}


\author[iu]{M.~D'Onofrio}

\author[uwcenpa]{A.~Eberhardt}

\author[ut]{Y.~Efremenko}

\author[lanl]{S.R.~Elliott}

\author[kurchatov,mephi]{A. Etenko}

\author[ornl]{L.~Fabris}

\author[ornl]{M.~Febbraro}

\author[efi,kicp,uofcphys]{N.~Fields}

\author[iu]{W.~Fox}

\author[uwcenpa]{Z.~Fu}

\author[ut,ornl]{A.~Galindo-Uribarri}

\author[tunl,ornl,ncsu]{M.P.~Green}

\author[efi,kicp,uofcphys]{M.~Hai}


\author[iu]{M.R.~Heath}

\author[tunl,duke]{S.~Hedges}

\author[ornl]{D.~Hornback}

\author[pnnl]{T.W.~Hossbach}

\author[ornl]{E.B.~Iverson}

\author[nmsu]{M.~Kaemingk}

\author[iu]{L.J.~Kaufman\fnref{fnslac}}
\fntext[fnslac]{Also at SLAC National Accelerator Laboratory, Menlo Park, CA 94205, USA}

\author[lbnl]{S.R.~Klein}

\author[mephi]{A.~Khromov}

\author[tunl,duke]{S.~Ki}

\author[kurchatov,mephi,itep]{A.~Konovalov}

\author[kurchatov,mephi]{A.~Kovalenko}

\author[tunl]{M.~Kremer}

\author[mephi]{A.~Kumpan}

\author[tunl]{C.~Leadbetter}

\author[tunl,duke]{L.~Li}

\author[ornl]{W.~Lu}

\author[tunl,ncsu]{K.~Mann}

\author[nccu,tunl]{D.M.~Markoff}

\author[mephi]{Y.~Melikyan}

\author[kicp,uofcphys]{K.~Miller}

\author[nmsu]{H.~Moreno}


\author[ornl]{P.E.~Mueller}

\author[mephi]{P.~Naumov}

\author[ornl]{J.~Newby}

\author[pnnl]{J.L.~Orrell}

\author[pnnl]{C.T.~Overman}

\author[cmu]{D.S.~Parno}


\author[ornl]{S.~Penttila}

\author[efi,kicp,uofcphys]{G. Perumpilly}

\author[ornl]{D.C.~Radford}

\author[cmu]{R.~Rapp}

\author[ufl]{H.~Ray}

\author[duke,ornl]{J.~Raybern}

\author[sandia]{D.~Reyna}

\author[efi,kicp]{G.C.~Rich\fnref{oldAffil}\corref{corr}}
\ead{gcrich@uchicago.edu}
\cortext[corr]{Corresponding author}
\fntext[oldAffil]{Previously at Department of Physics and Astronomy, University of North Carolina at Chapel Hill, Chapel Hill, NC 27599, USA and Triangle Universities Nuclear Laboratory, Durham, North Carolina, 27708, USA}

\author[ufl]{D.~Rimal}

\author[kurchatov,mephi]{D.~Rudik}

\author[uwcenpa]{D.J.~Salvat}

\author[duke]{K. Scholberg} 

\author[efi,kicp,uofcphys]{B.~Scholz}

\author[duke]{G. Sinev}

\author[iu]{W.M.~Snow}

\author[mephi]{V.~Sosnovtsev}

\author[mephi]{A.~Shakirov}

\author[lbnl]{S.~Suchyta\fnref{rsl}}
\fntext[rsl]{Presently at Remote Sensing Laboratory, Joint Base Andrews, MD 20762, USA} 

\author[iu]{B.~Suh}

\author[iu]{R.~Tayloe}

\author[iu]{R.T.~Thornton}

\author[iu]{I.~Tolstukhin}


\author[iu]{J.~Vanderwerp}

\author[ornl]{R.L.~Varner}

\author[laurentian]{C.J.~Virtue}

\author[tunl]{Z.~Wan}

\author[kaist]{J.~Yoo}

\author[ornl]{C.-H.~Yu}

\author[tunl]{A.~Zawada}

\author[uwcenpa]{A.~Zderic}

\author[iu]{J.~Zettlemoyer}

%

\begin{abstract}
This release includes data and information necessary to perform independent analyses of the COHERENT result presented in Akimov \etal \cite{akimov2017}, arXiv:1708.01294 [nucl-ex].
Data is shared in a binned, text-based format, including both ``signal'' and ``background'' regions, so that counts and associated uncertainties can be quantitatively calculated for the purpose of separate analyses.
This document describes the included information and its format, offering some guidance on use of the data. 
Accompanying code examples show basic interaction with the data using Python.
\end{abstract}

\maketitle

\section{Overview}

\subsection{How to access this data release}
The data release, this document, and accompanying code examples are available in two locations: at \url{http://coherent.ornl.gov/data} and also on Zenodo (DOI: \href{http://dx.doi.org/10.5281/zenodo.1228631}{{\tt 10.5281/zenodo.1228631}}) with an accompanying Git repository available at \url{https://code.ornl.gov/COHERENT/codeExamples_dataRelease_april2018}.
See Sec. \ref{sec:citing} for how to cite this data.
Future releases from the COHERENT Collaboration will be shared in a similar fashion and will have release-specific documentation.

\subsection{Summary of release materials}
\begin{itemize}
	\item Data
	\begin{itemize}
		\item Coincidence region data, beam on
		\item Coincidence region data, beam off
		\item Anti-coincidence region data, beam on
		\item Anti-coincidence region data, beam off
	\end{itemize}
	
	\item SNS-related backgrounds
	\begin{itemize}
		\item Beam-exposure normalized PDF describing photoelectron distribution expected from prompt neutrons created by the Spallation Neutron Source (SNS)
	\end{itemize}
	
	\item Calibration
	\begin{itemize}
		\item ${}^{241}$Am light-yield calibration
		\item Quenching-factor measurement data points
	\end{itemize}
	
	\item Experimental parameters
	\begin{itemize}
		\item SNS beam exposure, distance to target, detector mass, etc.
		\item \csi acceptance efficiency
		\item Arrival-time distributions for signals associated with prompt, SNS neutrons and \cevns events from both the prompt and delayed neutrino populations
	\end{itemize}
\end{itemize}

\subsection{Select tabulated values}
In some cases, the information included here is represented by single values and associated uncertainty. 
For simplicity, some of these values are tabulated in Tab. \ref{tab:summary}, but are also included in a format similar to other data to enable automated reading of these values.
The complete list of ``single-value'' parameters are compiled in a YAML file as described in Sec. \ref{sec:singleValueParamFormat}.

\begin{table}
\centering
\begin{tabular}[h]{l l r}
	Parameter & Unit & Value \\
	\midrule
	Decay-at-rest neutrino production & $\nu$ / flavor / SNS proton  & $0.08 \pm 0.008$ \\
	SNS beam exposure & GW-hr & 7.47594\\
	\csi quenching factor & \% & $8.78 \pm 1.66$ \\
	\csi light yield at 59.54 keVee & PE / keVee & $13.348 \pm 0.019$ \\
	\bottomrule
\end{tabular}
\caption[Tabulated parameters]{Limited sample of parameters which can be expressed as single values. Parameters such as these are contained in a YAML file included in the data release; an example Python script is included which reads this file and prints out the list of all included parameters (see Sections \protect{\ref{sec:singleValueParamFormat}} and \protect{\ref{sec:exampleCode}}). The quenching factor, a subject of considerable importance and the dominant uncertainty in the analysis of Akimov \etal \protect{\cite{akimov2017}}, was determined based on COHERENT measurements in addition to literature values (see supplemental materials of Ref. \protect{\cite{akimov2017}}).}
\label{tab:summary}
\end{table}



\section{Detailed description of the release}
This release includes information which can be broadly sorted into the following categories:
\begin{description}
	\item[Data] The actual data, representing both SNS beam-on and beam-off periods and including, in both cases, the coincidence and anti-coincidence regions of the waveform (discussed below; see also Fig. \ref{fig:csiWaveform})
	\item[SNS-related backgrounds] PDF describing the contributions expected from promptly arriving, SNS-produced neutrons
	\item[Calibration] Information relating to the signal yield expected from the \csi detector in electron-equivalent energy and information related to the translation from nuclear-recoil energy into electron-equivalent energy
	\item[Experimental parameters] Information related to SNS operation, neutrino production, and signal timing, including timing distributions for beam-related signal components
\end{description}

\subsection{Data}
The data is presented as a binned, two-dimensional histogram, representing the arrival time of the signal in the \csi detector and the number of photoelectrons present in the signal.
In correspondence with the finest binning presented in Akimov \etal \cite{akimov2017}, the bin widths are 0.5 \us and 2 PE for the arrival time and photoelectron axes, respectively.

\begin{figure}
	\centering
	\includegraphics[width=0.7\linewidth]{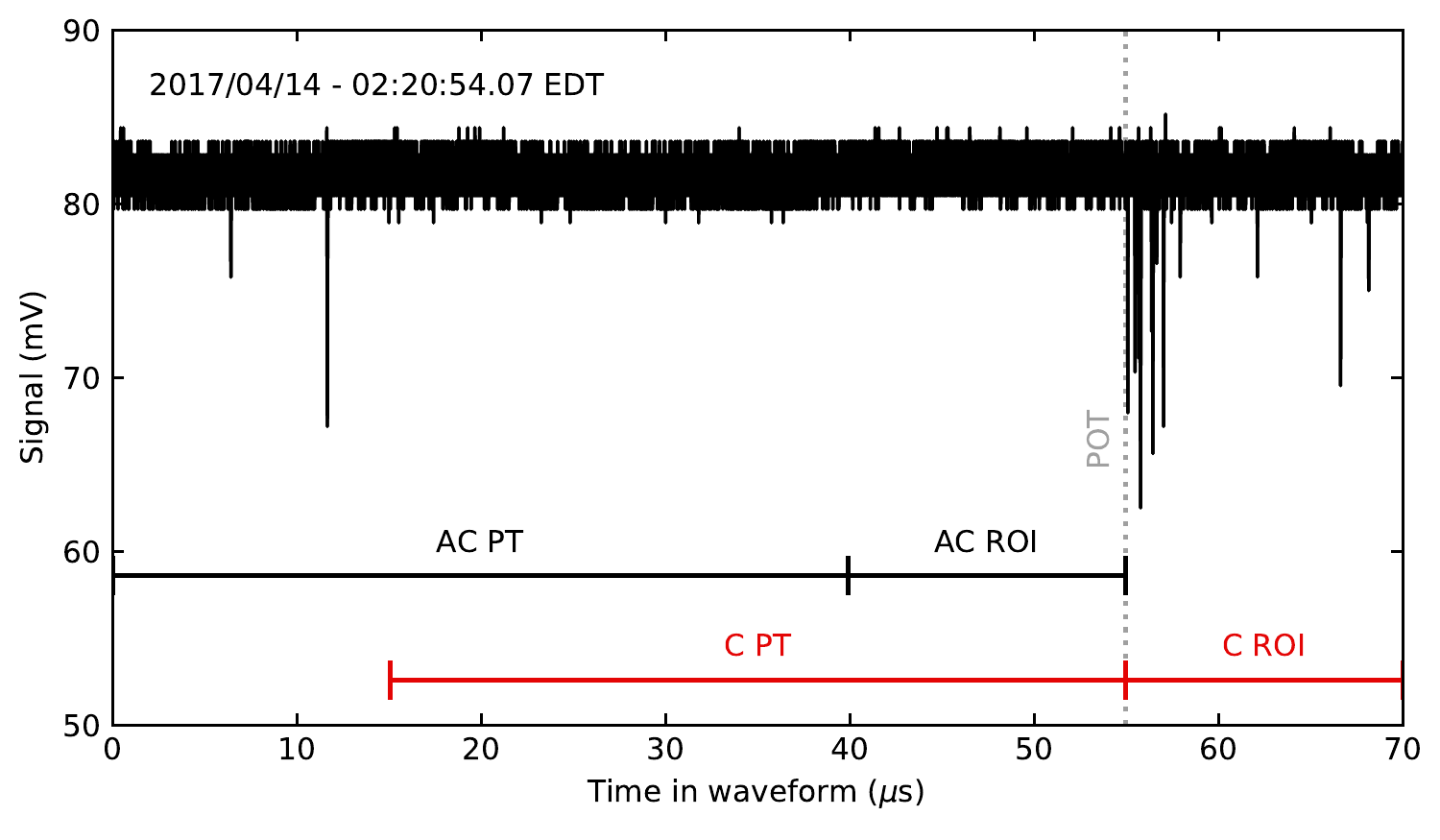}
	\caption[Example waveform from \cevns search, showing C and AC regions]{An example waveform from the \csi detector acquired during the \cevns search at the SNS. The coincidence (C) and anti-coincidence (AC) regions are highlighted, as well as the time at which the SNS timing signal triggered the DAQ system (labeled POT). The C and AC regions both include a pretrace (PT) and region-of-interest (ROI). Information from the pretrace is not included in this release, but is encoded in the included signal acceptance efficiency. Figure from Reference \protect{\cite{scholzThesis}}.}
	\label{fig:csiWaveform}
\end{figure}

For each trigger to the DAQ, the waveform is divided into two equal-length regions, described here:
\begin{description}
	\item[Coincidence (C)] The region of the waveform where signals associated with the SNS beam (both neutrinos as well as neutrons) are expected; could be referred to as ``signal'' region
	\item[Anti-coincidence (AC)] A region of the waveform where \emph{no contribution from the SNS beam is expected}; could be considered a ``background'' region
\end{description}
Data for each of these regions is included in this release; Fig. \ref{fig:csiWaveform} shows a waveform acquired during the \cevns search and graphically depicts the C and AC regions.
Both the C and AC regions have an accompanying pretrace region (denoted PT in Fig. \ref{fig:csiWaveform}); these pretraces are used to reject events which do not pass certain quality cuts, and may for instance show contamination from afterglow associated with a preceding event lying outside of the waveform.
The effect of these pretrace cuts, in addition to other data quality cuts, are included in this data release via the acceptance efficiency Eq. \eqref{eq:acceptance} and its parameter values.

The SNS continues to provide timing signals during periods when no beam is incident upon the target. 
Consequently, data was collected using the same trigger configuration during both \textbf{beam-on} and \textbf{beam-off} periods.
Data from both of these periods is included, representing both the coincidence and anti-coincidence regions.


\subsection{SNS backgrounds}
As discussed in Akimov \etal \cite{akimov2017}, there are several background features that must be accounted for when analyzing the \cevns search data.
The beam-on, AC region data informs a model for the steady-state background component (see Refs. \cite{akimov2017,richThesis} and Sec. \ref{sec:backgroundPDF}), but another component which must be accounted for is that associated with neutrons produced by the SNS.

Making use of additional measurements at the SNS, a probability distribution function describing the expected recoil-energy distribution associated with prompt SNS neutrons was produced; this PDF is included in the data release.
The methods by which this PDF was derived are outside the scope of this discussion; for details, see the supplementary materials of Ref. \cite{akimov2017}.

This prompt-neutron PDF is included as a beam-exposure normalized distribution in photoelectron space (i.e., it is in units of Counts / bin / GW-hr) with 1-PE bin widths.
The binning is slightly different from that of the data included in this release; the finer binning of the prompt-neutron distribution allows application of the analysis acceptance described in Sec. \ref{sec:expParams} in a manner consistent with the analysis of Akimov \etal \cite{akimov2017}.
The timing distribution of these events is also included and discussed in Sec. \ref{sec:neutrinoEnergyAndTime}. 


\subsection{Calibration}
Calibration information is included which describes the response of the \csi detector to incident neutrons and $\gamma$-rays.
An ${}^{241}$Am source was used to determine the light yield of the \csi detector at the full-energy peak of 59.54 keV (see Ref. \cite{scholzThesis} or the supplementary materials of Ref. \cite{akimov2017} for more details about these calibrations).
In the analysis of Ref. \cite{akimov2017}, the yield at this data point was extrapolated linearly to 0, providing the functional light-yield estimate. 

The ${}^{241}$Am calibration describes the light yield in terms of electron-equivalent energies. 
\cevns signals are produced by \emph{nuclear recoils}, so a conversion from keVnr (nuclear recoil energy) to keVee (electron equivalent energy) must be made.
This is facilitated by the quenching factor, which was measured by two experiments conducted within the COHERENT Collaboration (see Refs. \cite{scholzThesis,richThesis} for extensive discussion).

Analysis presented in Akimov \etal \cite{akimov2017} was based on an energy independent treatment of the QF. 
We include here the value used for the previous analysis and also include the data points measured by the COHERENT groups, allowing for alternative treatments of QF parameterization.

\begin{figure}
	\centering
	\includegraphics[width=0.7\linewidth]{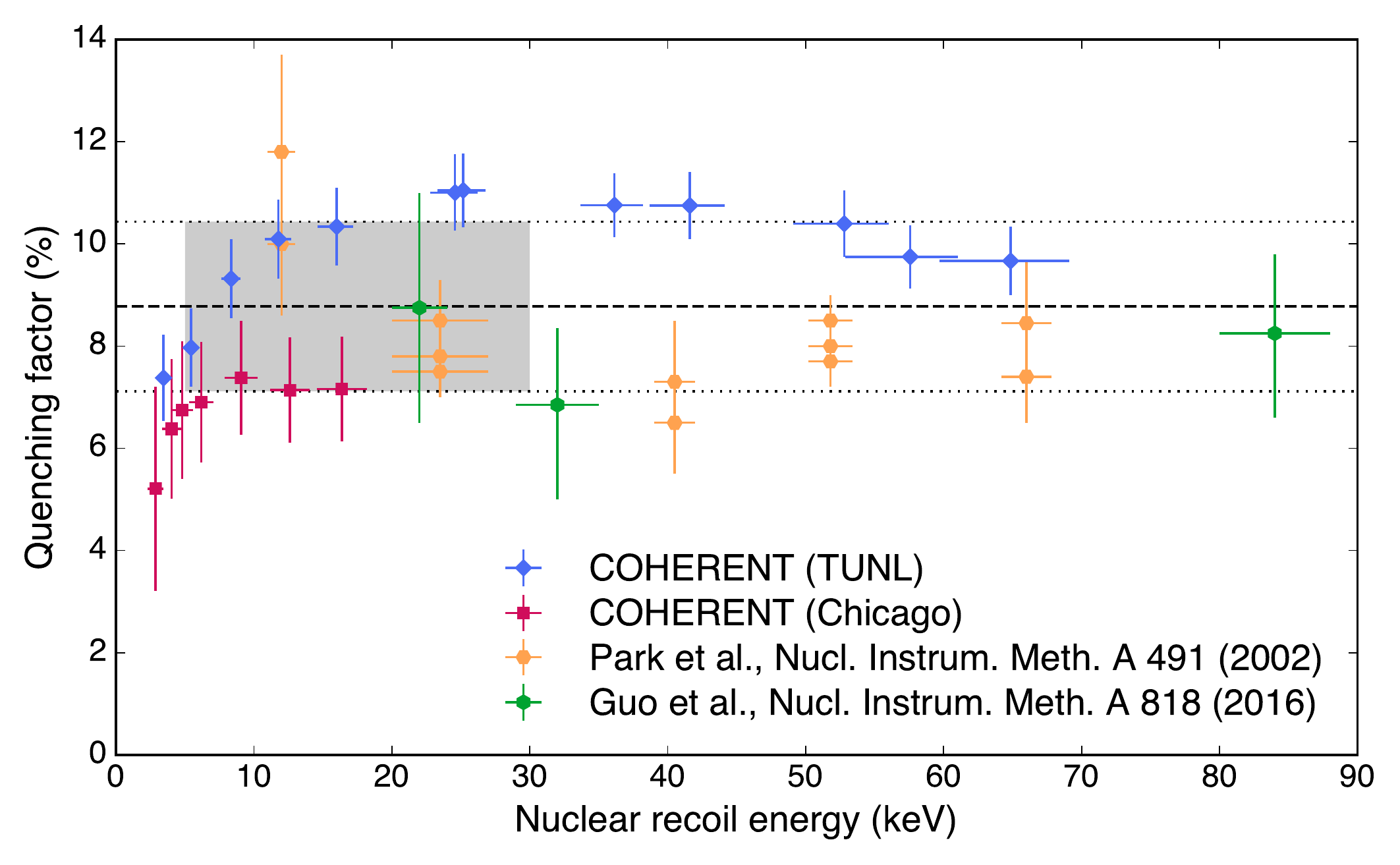}
	\caption[Global \csi quenching factor data]{Quenching factor data for \csi. The values from the COHERENT Collaboration are included in this release. Data from Park \etal \protect{\cite{park2002csiQF}} and Guo \etal \protect{\cite{guo2016csiQF}} was extracted from the original works. Figure appears in Reference \protect{\cite{richThesis}}, adapted from Akimov \etal \protect{\cite{akimov2017}}.}
	\label{fig:qf}
\end{figure}


\subsection{Experimental parameters} \label{sec:expParams}

\begin{figure}
	\centering
	\includegraphics[width=0.5\linewidth]{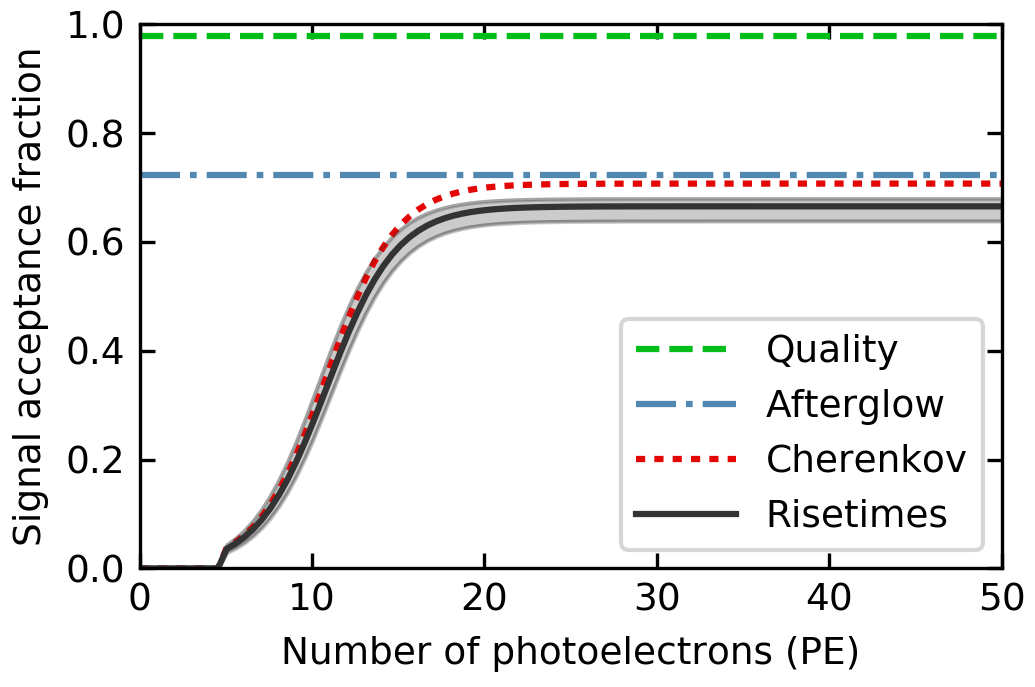}
	\caption[Acceptance efficiency curve]{Acceptance efficiency curve for \csi analysis described by Equation \protect{\eqref{eq:acceptance}}. From supplementary materials of Akimov \etal \cite{akimov2017}.}
	\label{fig:acceptance}
\end{figure}

The processing and analysis of the \csi data imposed an acceptance efficiency in terms of the photoelectron content of the signal $x$ which can be described by
\begin{equation} \label{eq:acceptance}
	f\args{x} = \frac{a}{1 + \exp{\left( -k \left( x-  x_0\right)\right)}} \Theta\args{x-5},
\end{equation}
where $\Theta \args{x}$ is a modified Heaviside step function and the parameters have values \cite{scholzThesis}
\begin{align*}
	a &= 0.6655 ^{+0.0212}_{-0.0384}, \\
	k &= 0.4942 ^{+0.0335}_{-0.0131}, \\
	x_0 &= 10.8507 ^{+0.1838}_{-0.3995},
\end{align*}
and are treated as uncorrelated.
The function $\Theta \args{x}$ is defined as 
\[
	\Theta \args{x} = \begin{cases}
		0 & x < 5, \\
		0.5 & 5 \leq x < 6, \\
		1 & x \geq 6.
	\end{cases}
\]
Figure \ref{fig:acceptance} shows this curve and a Python implementation of the efficiency is offered as function {\tt csiEfficiency} in the example code, specifically in file {\tt coherent\_readPromptPDF.py}.
This acceptance function was applied to \emph{binned} models in photoelectron space with a bin width of 1 PE; for each bin, the effective acceptance was taken as the value of Eq. \eqref{eq:acceptance} at the bin center (e.g., 0.5 PE for 1 PE signals, 1.5 PE for 2 PE signals, etc).

Details such as distance to target, target mass, and information needed to calculate the neutrino flux are all included in this data release.
For a full list of the included parameters, use the included Python script {\tt coherent\_readParameters.py} to parse the included YAML file which contains this information; this will print out a list of the parameters in the collection.

\subsubsection{Energy and time distributions of SNS neutrinos} \label{sec:neutrinoEnergyAndTime}

\begin{figure}
	\centering
	\includegraphics[width=0.7\linewidth]{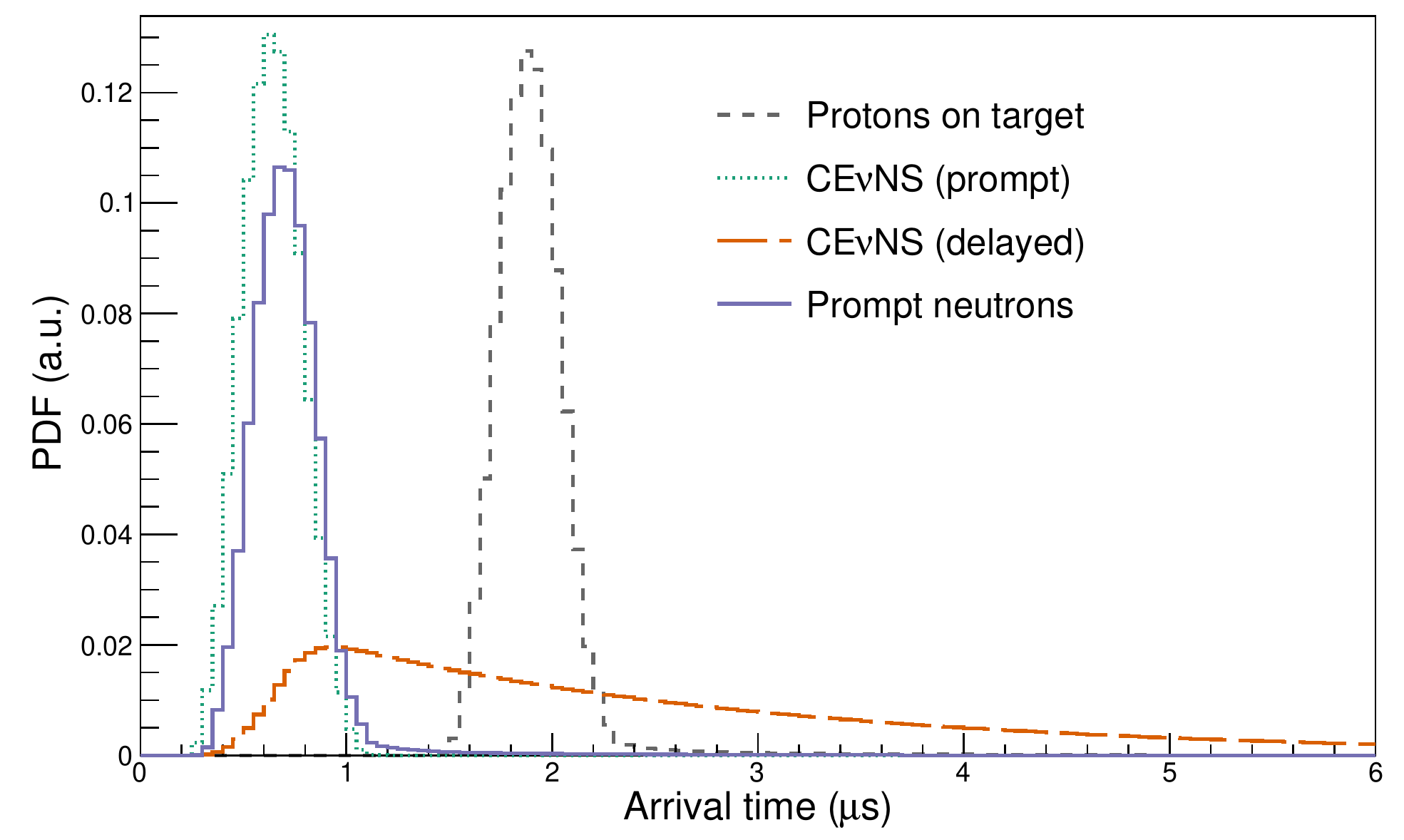}
	\caption[Neutrino interaction timing distribution at the SNS]{Timing distributions for neutrino interactions at the CsI[Na] detector at the SNS. The prompt neutrino component, drawn as a dotted, teal line, is associated with interactions from the \protect{$\nu_\mu$} component of the neutrino population, which arises from the decay of stopped pions in the mercury target of the SNS; this time distribution is produced by convolving the protons-on-target (POT) PDF shown coarsely in a dashed gray line with an exponential decay of \protect{$\tau = 26~\textrm{ns}$}, the time constant for pion decay. Muons are produced in the pion decay and the decay of these muons, a high fraction of which are stopped in the mercury target, yields a delayed neutrino population which is shown by a dashed, burnt-orange line. The timing distribution of the delayed neutrino population is produced via convolution of the \emph{prompt}-neutrino distribution, representing muon production from pion decay, with a 2.2-\protect{$\mu$}s exponential representing the decay of those muons. Due to processing and propagation delays of the POT signal, it is arbitrarily offset from signals in the \csi; this offset was experimentally determined by an \emph{in situ} measurement of neutron signals (see Section \protect{\ref{sec:neutrinoEnergyAndTime}}). From Reference \protect{\cite{richThesis}}.}
	\label{fig:snsTiming}
\end{figure}

Analyses of the \cevns data can be strengthened by use of both energy and timing information.
The distribution in time of SNS protons arriving on the target was extracted from the SNS diagnostics system and an average of the distribution observed over the course of the data acquisition period was used to inform distributions of beam-related signal components.
Distributions in arrival-time space for the prompt-neutron background along with the \cevns populations, both prompt and delayed, are included in this release.
Examples of these timing distributions can be seen in Fig. \ref{fig:snsTiming}.

In the analysis of Ref. \cite{akimov2017}, the energy and timing distributions of the spectral components were treated as uncorrelated.
Making this (good) approximation, the arrival-time PDFs can be used to project energy-space PDFs into arrival-time space as well\footnote{Note that normalization must be considered when performing this ``extension'' into the time dimension. Most \cevns calculations would likely predict the distribution in recoil energy, so the energy distribution would carry the proper normalization; this is the case for the prompt neutron PDF included here, as well. Consequently, the time-space PDF used to perform this operation should be normalized to unity, which is done for the included time-space PDFs.}. 


It is important to note that the timing of the POT signal is effectively arbitrary by the time it arrives at the \csi data acquisition system, triggering the digitizer.
Using data from a deployment of liquid scintillator cells to the same location at the SNS as the \csi was eventually installed, the relative offset between prompt neutrons and the timing signal was determined.
The relative time between neutrino and prompt-neutron signals was taken from GEANT4 simulations and was in agreement with the value extracted from the liquid scintillator cell data.
See a discussion of the \emph{in situ} neutron measurement in the Supplementary Materials of Ref. \cite{akimov2017}.
Due to coarse time binning in the analysis, uncertainties on the timing offsets were excluded from consideration in Ref. \cite{akimov2017} and they are correspondingly not included here.

\begin{figure}
	\centering
	\includegraphics[width=0.5\linewidth]{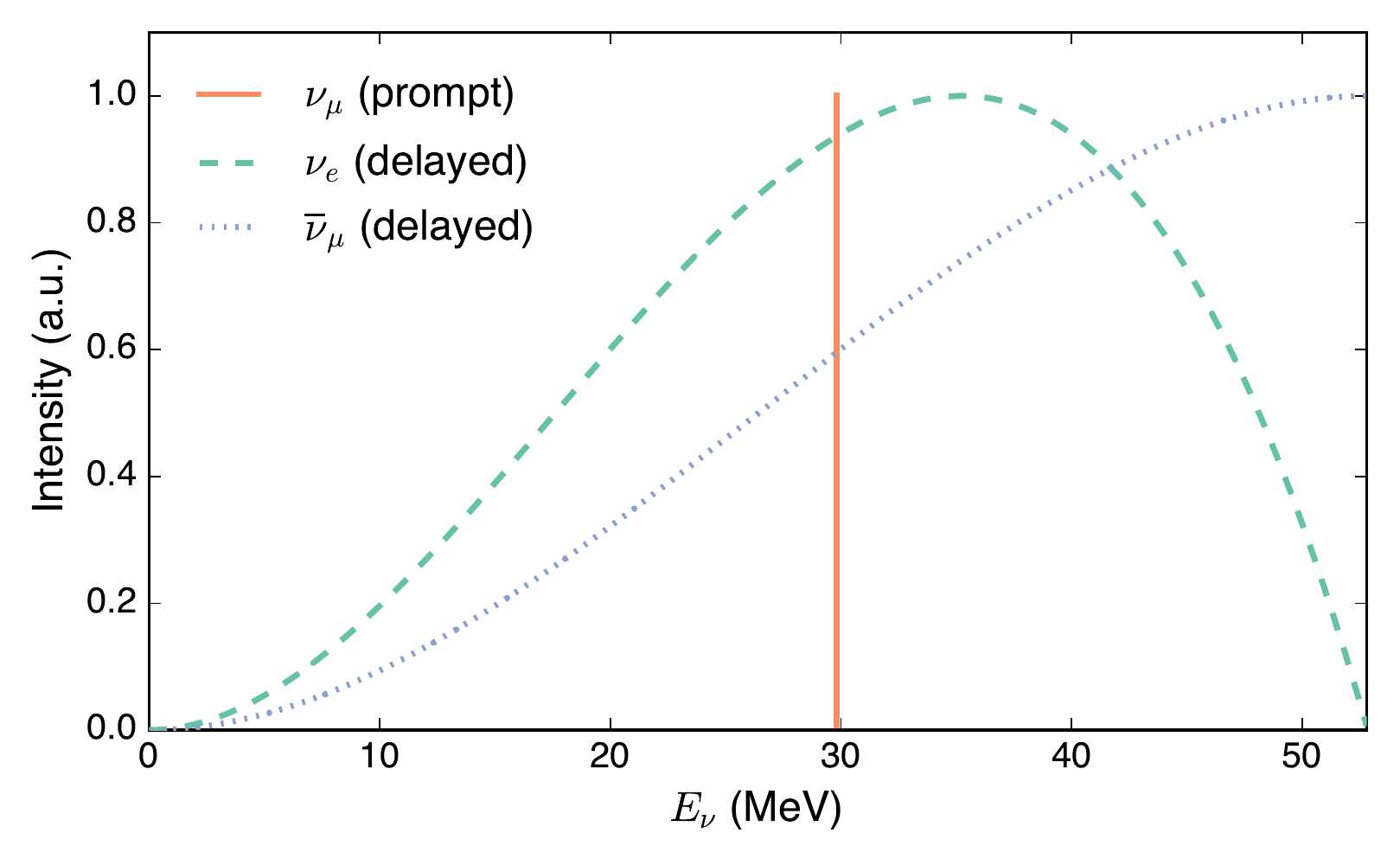}
	\caption[Approximate neutrino energy distributions from stopped pion neutrinos]{Energy distributions for stopped-pion neutrino sources like the SNS. The delayed neutrino population has an energy distribution described by the Michel spectrum, while the prompt \numu neutrinos are monoenergetic with \protect{$E_\numu = 29.8$} MeV. As this well approximates the results of the GEANT4 simulation, use of this simple description is recommended. Image from Reference \protect{\cite{richThesis}}.}
	\label{fig:snsNuEnergy}
\end{figure}

Results of GEANT4 simulations for the neutrino energy distribution are shown in the supplementary materials of Ref. \cite{akimov2017}, but they are well approximated by the standard treatment for stopped pion neutrino sources: a Michel spectrum describing the energy of the delayed population of \numubar and \nue neutrinos and a monoenergetic population of prompt \numu neutrinos with $E_\numu = 29.8$ MeV.
This distribution is described many places in the literature, including e.g. Ref. \cite{amanik2008}, and is seen in Fig. \ref{fig:snsNuEnergy}.

\section{Format of data} \label{sec:format}


\subsection{Data}
The data is saved in a comma-separated value (CSV)-format text file.
There are 3 columns in each file, and each file has some header information denoted by lines beginning with `\#'.
The columns, also described in the header, represent: the central value of the bin in photoelectron space; the central value of the bin in arrival time space; and the number of events \emph{in this bin} (there is no adjustment of this value for bin width).
Table \ref{tab:dataFilenames} relates the data filenames to the data collection stored within.

\begin{table}
\centering
\begin{tabular}[h]{l l l}
	Description & Comment & Filename \\
	\midrule
	\multirow{4}{*}{Data} & SNS beam on, coinc. region & {\tt data\_coincidence\_beamOn.txt}  \\
	& SNS beam on, anti-coinc. region & {\tt data\_anticoincidence\_beamOn.txt}  \\
	& SNS beam off, coinc. region & {\tt data\_coincidence\_beamOff.txt} \\
	& SNS beam off, anti-coinc. region & {\tt data\_anticoincidence\_beamOff.txt} \\
	\multirow{2}{*}{Quenching factors} & TUNL experiment & {\tt qfData\_tunl.txt} \\
	& Chicago experiment & {\tt qfData\_chicago.txt} \\
	Prompt neutron PDF & & {\tt promptPDF.txt} \\
	\multirow{3}{*}{Arrival-time distributions} & {Prompt neutrons} & {\tt arrivalTimePDF\_promptNeutrons.txt}\\
	 & {Prompt neutrinos} & {\tt arrivalTimePDF\_promptNeutrinos.txt}\\
	 & {Delayed neutrinos} & {\tt arrivalTimePDF\_delayedNeutrinos.txt}\\
	Other parameters & YAML format & {\tt coherent\_parameters.yaml}\\
	\bottomrule
\end{tabular}
\caption[Filename descriptions]{Data filenames and the type of data stored within, specifying coincidence vs anti-coincidence waveform region and whether or not the SNS beam was active during acquisition. All of these files are contained within the ``{\tt data}'' directory of the release. All are plaintext files and most are CSV formatted with the exception of parameters which can be expressed as single values: these are contained in a YAML file. Section \protect{\ref{sec:format}} of this text discusses the data formats, and for details of the example Python code included in this release which reads these files, see Section \protect{\ref{sec:exampleCode}}.}
\label{tab:dataFilenames}
\end{table}


\subsection{Calibration, experimental parameters, and distributions} \label{sec:singleValueParamFormat}
To compactly store the relevant parameters which can be expressed by a single number with some uncertainty, a YAML file is included which contains entries pertaining to these data points.
The format of this file is intended to be easily interpretable, with comments on the parameters included in the YAML file. 
To assist in the handling of this file, a Python script is included that demonstrates accessing the entries and retrieving their values.

YAML is a text-based, human-readable format. 
The presentation of the parameters stored in this file is intended to be readily interpretable.
An example entry is here:
\begin{keeptogether}
\begin{lstlisting}[language=yaml]
neutrinosPerProton:
  name: Neutrinos produced per SNS proton
  value: 0.08
  uncertainty: 0.008
  units: nu/flavor/proton
  comment: |
    This is the number of decay-at-rest neutrinos produced per incident SNS beam proton per neutrino flavor.
    The DAR neutrinos are emitted isotropically.
\end{lstlisting}
\end{keeptogether}

\subsubsection{Quenching factor data points}
The quenching factor data points for the TUNL and Chicago experiments are stored in separate text-based CSV files, with a unique file for each experiment.
Header lines begin with `\#' and describe the ordering of the columns.
For each experiment, data points are specified by their mean recoil energy and the measured quenching factor at that recoil energy; symmetric uncertainties for both recoil energy and quenching factor are included for the Chicago data, while asymmetric uncertainties are included for the TUNL data. 
The horizontal uncertainties represent the spread of nuclear-recoil energies tagged for each energy point.
Note that in this treatment, the horizontal and vertical errors are correlated in a non-trivial way. 
In the simple treatment of Akimov \etal \cite{akimov2017}, this effect was conservatively approximated by redefining the error on the quenching factor as the sum in quadrature of the independently determined fractional errors on the recoil energy and the quenching factor.

\subsubsection{Prompt-neutron signal PDF and timing distributions}
Similar to the format for the data, the PDF describing the PE distribution from prompt neutrons is saved in a plain-text CSV file.
Column headers, on `\#'-specified lines, describe the column contents, which represent the bin center (in number of photoelectrons) and the bin contents.
The timing distributions for the prompt-neutron signal and both the prompt and delayed \cevns populations are included in the same format: a CSV file with a `\#'-specified header describing the columns which represent the bin centers (in arrival time) and the bin contents (in fractional intensity).

A few comments on these distributions
\begin{itemize}
	\item As included, the prompt neutron distribution \emph{does not} have the analysis efficiency applied, and the efficiency described by Eq. \eqref{eq:acceptance} must be incorporated prior to use with data; the included Python code addresses this, see Sec. \ref{sec:exampleCode}.
	\item Timing distributions are normalized to unity. When constructing 2-D PDFs in the analysis of Akimov \etal \cite{akimov2017}, these timing distributions were multiplied with beam-exposure normalized PDFs describing expected count rates in terms of photoelectrons.
\end{itemize}


\section{Example code} \label{sec:exampleCode}
\begin{figure}[h]
	\centering
	\begin{subfigure}[h]{0.45\textwidth}
		\includegraphics[width=\textwidth]{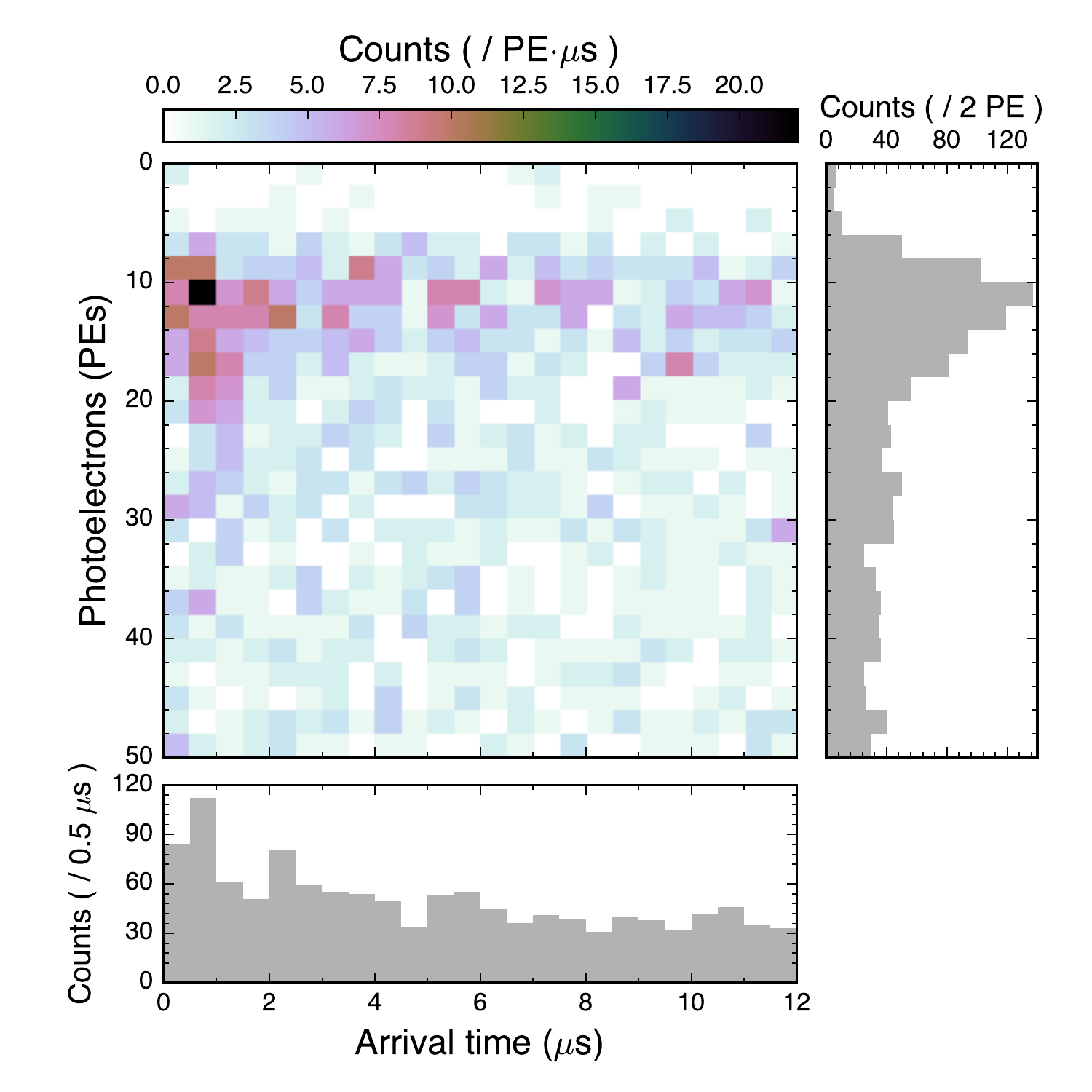}
		\caption[Example 2-D histogram of coincidence, beam-on data]{Image produced by packaged script {\tt coherent\_readDataExample.py}}
		\label{fig:example2D}
	\end{subfigure}
	~
	\begin{subfigure}[h]{0.45\textwidth}
		\includegraphics[width=\textwidth]{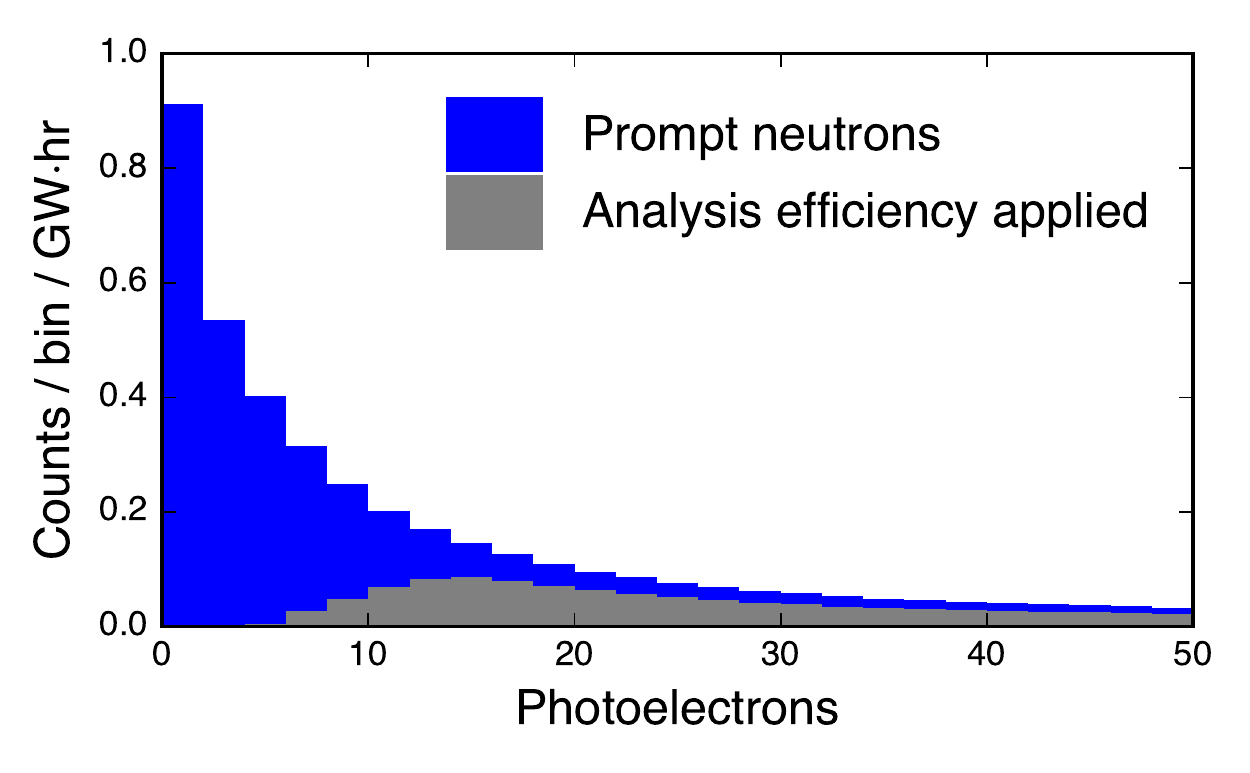}
		\caption{Image from {\tt coherent\_readPromptPdf.py}}
		\label{fig:examplePromptPDF}
	\end{subfigure}
	~
	\begin{subfigure}[h]{0.45\textwidth}
		\includegraphics[width=\textwidth]{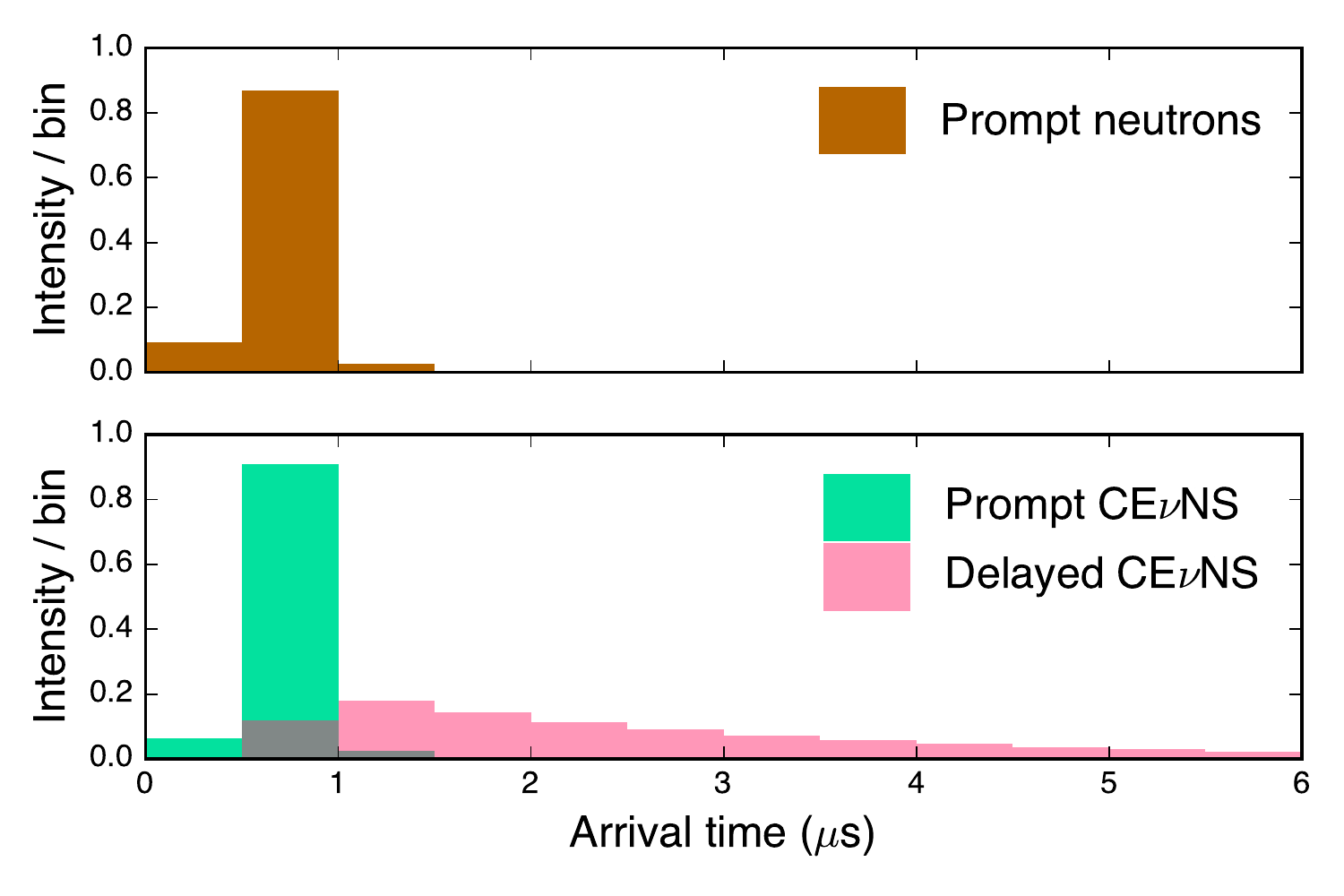}
		\caption{Image of timing distributions for both neutron and neutrino components of the signal, produced by {\tt coherent\_readTiming.py}}
		\label{fig:timingDistributions}
	\end{subfigure}
	
	\caption[Example images from included Python scripts]{Images produced with the included Python scripts. Plot (a) shows the 2-D data from the beam-on, coincidence region; (b) is the expected contribution per GW-hr exposure from prompt SNS neutrons; and (c) shows the timing distributions for the prompt-neutron background contribution as well as both the prompt and delayed \cevns components, restricted to the first 6 $\mu$s. In all cases, the bin contents are absolute and have no multiplicative factor associated with bin width. See a discussion of the timing distributions in the caption of Figure \protect{\ref{fig:snsTiming}} and the text of Section \protect{\ref{sec:neutrinoEnergyAndTime}}.}
	\label{fig:exampleImages}
\end{figure}

Included in this release, in the ``{\tt code}'' directory, are some example Python scripts which make basic use of the data.
Many of these scripts produce images which can be compared with those shown in Fig. \ref{fig:exampleImages}.

\begin{description}
	\item[{\tt coherent\_readDataExample.py}] produces an image like that shown in Fig. \ref{fig:example2D}, depicting the 2-D histogram built from the coincidence-region, beam-on data
	\item[{\tt coherent\_readPromptPDF.py}] parses the included PDF describing the photoelectron distribution expected from prompt SNS neutrons; it produces an image overlaying the raw distribution with the expected distribution in the \csi data, having applied the efficiency cut of Eq. \eqref{eq:acceptance}, as can be seen in Fig. \ref{fig:examplePromptPDF}
	\item[{\tt coherent\_readParameters.py}] reads the included YAML file which contains relevant parameters that can be expressed as single numbers
	\item[{\tt coherent\_readTiming.py}] reads the included timing distributions for signals from prompt neutrons and all neutrinos, producing a plot like that shown in Fig. \ref{fig:timingDistributions}
\end{description}

\section{Usage}

\subsection{Citing this release} \label{sec:citing}
If you make use of this data in your work, the COHERENT Collaboration requests that you cite both Akimov \etal \cite{akimov2017} in addition to the Zenodo posting of this dataset.
Suggested formatting for the Zenodo citation is (replacing {\tt xxxx.xxxxx} with the arXiv ID for this document): 
\begin{displayquote}
	D. Akimov \etal (2018). COHERENT Collaboration data release from the first observation of coherent, elastic neutrino-nucleus scattering [Data set]. Zenodo. DOI: 10.5281/zenodo.1228631. arXiv: {\tt xxxx.xxxxx [nucl-ex]}.
\end{displayquote}

\subsection{Development of steady-state background PDF} \label{sec:backgroundPDF}
The beam-on, AC data provides a sample of the steady-state background component that is expected to be present in the C-region (``signal'') data as well.
By treating this data as uncorrelated in energy and time, it is possible to avoid some issues that may be introduced by the limited statistics of the sample.
In Akimov \etal \cite{akimov2017}, the AC data was projected onto the time axis and an exponential was fit to this projection; the exponential shape is used as an effective model as this background component is expected to represent the arrival time of accidental signals.
This exponential, normalized to unity, was then multiplied by the AC data projected onto the photoelectron axis to produce a 2-D PDF representing the accidental, steady-state background component. 
The overall normalization of this component was constrained by a Poisson distribution with $\mu$ equal to the number of counts in the AC-region data.

\subsection{General comments on use}
Robust independent analysis of this data of course requires many steps which diverge from the basic examples provided in the code accompanying this release.
A generic outline may be as follows
\begin{enumerate}
	\item Perform a calculation of the expected \emph{nuclear recoil} distribution from \cevns, describing a shape in keVnr, using e.g., Equation 11 in Ref. \cite{barranco2007}
	\item Utilize knowledge of the quenching factor for nuclear recoils in \csi (measurements of which are included here) to convert the distribution into electron-equivalent energy (keVee)
	\item Convert the electron-equivalent energy distribution into a distribution in photoelectrons, using the ${}^{241}$Am light-yield calibration included in this release
	\item For the predicted \cevns and prompt-neutron distributions in PE space, binned with bin widths of 1 PE, apply the acceptance described by Eq. \eqref{eq:acceptance}
	\item Compare the acceptance-applied, predicted PE distribution from \cevns, summed appropriately with background models representing the steady-state and prompt-neutron components, with the observed data, calculating a likelihood or $\chi^2$
	\item Adjust floating parameters in \cevns cross-section model and iterate to minimize/maximize chosen criterion
\end{enumerate}
In performing an analysis like that outlined here, one must make use of the calibrations and associated uncertainties included in this release to produce appropriate distributions in photoelectron-space, including uncertainties, which may then be compared with the data.


\bibliographystyle{vitae}
\bibliography{dataReleaseCompanion}

\begin{thebibliography}{1}
\expandafter\ifx\csname url\endcsname\relax
  \def\url#1{\texttt{#1}}\fi
\expandafter\ifx\csname urlprefix\endcsname\relax\def\urlprefix{URL }\fi
\providecommand{\bibAnnoteFile}[1]{%
  \IfFileExists{#1}{\begin{quotation}\noindent\textsc{Key:} #1\\
  \textsc{Annotation:}\ \input{#1}\end{quotation}}{}}
\providecommand{\bibAnnote}[2]{%
  \begin{quotation}\noindent\textsc{Key:} #1\\
  \textsc{Annotation:}\ #2\end{quotation}}
\providecommand{\bibinfo}[2]{#2}
\providecommand{\eprint}[2][]{\url{#2}}

\bibitem{akimov2017}
\bibinfo{author}{D.~Akimov} \emph{et~al.}
\newblock \enquote{\bibinfo{title}{{Observation of coherent elastic
  neutrino-nucleus scattering}}.}
\newblock \emph{\bibinfo{journal}{Science}} \textbf{\bibinfo{volume}{357}},
  \bibinfo{pages}{1123--1126} (\bibinfo{year}{2017}).
\newblock \eprint{1708.01294}.
\bibAnnoteFile{akimov2017}

\bibitem{scholzThesis}
\bibinfo{author}{B.J. Scholz}.
\newblock \emph{\bibinfo{title}{{First observation of coherent elastic
  neutrino-nucleus scattering}}}.
\newblock Ph.D. thesis, \bibinfo{school}{{University of Chicago}}
  (\bibinfo{year}{2017}).
\bibAnnoteFile{scholzThesis}

\bibitem{richThesis}
\bibinfo{author}{G.C. Rich}.
\newblock \emph{\bibinfo{title}{{Measurement of low-energy nuclear-recoil
  quenching factors in CsI{[Na]} and statistical analysis of the first
  observation of coherent, elastic neutrino-nucleus scattering}}}.
\newblock Ph.D. thesis, \bibinfo{school}{{University of North Carolina at
  Chapel Hill}} (\bibinfo{year}{2017}).
\bibAnnoteFile{richThesis}

\bibitem{park2002csiQF}
\bibinfo{author}{H.~Park}, \bibinfo{author}{D.H. Choi}, \bibinfo{author}{J.M.
  Choi}, \bibinfo{author}{I.S. Hahn}, \bibinfo{author}{M.J. Hwang},
  \bibinfo{author}{W.G. Kang}, \bibinfo{author}{H.J. Kim},
  \bibinfo{author}{J.H. Kim}, \bibinfo{author}{S.C. Kim}, \bibinfo{author}{S.K.
  Kim}, \emph{et~al.}
\newblock \enquote{\bibinfo{title}{Neutron beam test of CsI crystal for dark
  matter search}.}
\newblock \emph{\bibinfo{journal}{Nucl. Instrum. Meth. A}}
  \textbf{\bibinfo{volume}{491}}, \bibinfo{pages}{460--469}
  (\bibinfo{year}{2002}).
\bibAnnoteFile{park2002csiQF}

\bibitem{guo2016csiQF}
\bibinfo{author}{C.~Guo}, \bibinfo{author}{X.H. Ma}, \bibinfo{author}{Z.M.
  Wang}, \bibinfo{author}{J.~Bao}, \bibinfo{author}{C.J. Dai},
  \bibinfo{author}{M.Y. Guan}, \bibinfo{author}{J.C. Liu},
  \bibinfo{author}{Z.H. Li}, \bibinfo{author}{J.~Ren}, \bibinfo{author}{X.C.
  Ruan}, \emph{et~al.}
\newblock \enquote{\bibinfo{title}{Neutron beam tests of CsI(Na) and
  CaF${}_2$(Eu) crystals for dark matter direct search}.}
\newblock \emph{\bibinfo{journal}{Nucl. Instrum. Meth. A}}
  \textbf{\bibinfo{volume}{818}}, \bibinfo{pages}{38--44}
  (\bibinfo{year}{2016}).
\bibAnnoteFile{guo2016csiQF}

\bibitem{amanik2008}
\bibinfo{author}{P.S. Amanik} and \bibinfo{author}{G.C. McLaughlin}.
\newblock \enquote{\bibinfo{title}{Nuclear neutron form factor from
  neutrino--nucleus coherent elastic scattering}.}
\newblock \emph{\bibinfo{journal}{J. Phys. G Nucl. Partic.}}
  \textbf{\bibinfo{volume}{36}}, \bibinfo{pages}{015105}
  (\bibinfo{year}{2008}).
\bibAnnoteFile{amanik2008}

\bibitem{barranco2007}
\bibinfo{author}{J.~Barranco}, \bibinfo{author}{O.G. Miranda}, and
  \bibinfo{author}{T.I. Rashba}.
\newblock \enquote{\bibinfo{title}{Sensitivity of low energy neutrino
  experiments to physics beyond the standard model}.}
\newblock \emph{\bibinfo{journal}{Phys. Rev. D}} \textbf{\bibinfo{volume}{76}},
  \bibinfo{pages}{073008} (\bibinfo{year}{2007}).
\bibAnnoteFile{barranco2007}

\end{thebibliography}

\end{document}